\documentclass[a4paper]{article}
\usepackage{spconf,amsmath,graphicx,cite,verbatim,caption}
\usepackage[utf8]{inputenc}
\usepackage{color}
\usepackage{balance}
\title{Binaural coherent-to-diffuse-ratio estimation \\ for dereverberation using an ITD model}
\name{\normalsize Chengshi Zheng$^*$$^\dagger$, Andreas Schwarz$^\dagger$, Walter Kellermann$^\dagger$, Xiaodong Li$^*$ %
   \thanks{This work was supported by the National Science Fund of
China (NSFC) under Grants 61201403 and 61302126.}%
}
\address{%
    \tabular{c}
       $^*$ \normalsize Communication Acoustics Laboratory\\
       \normalsize Institute of Acoustics, CAS\\
       \normalsize 100190 Beijing, China
   \endtabular
   \hskip 0.5in
    \tabular{c}
       $^\dagger$ \normalsize Chair of Multimedia Communications and Signal Processing \\
\normalsize Friedrich-Alexander-Universit\"at Erlangen-N\"urnberg\\
\normalsize 91058 Erlangen, Germany
   \endtabular
}
\begin{document}

\maketitle
\begin{abstract}
Most previously proposed dual-channel coherent-to-diffuse-ratio
(CDR) estimators are based on a free-field model. When used for
binaural signals, e.g., for dereverberation in binaural hearing
aids, their performance may degrade due to the influence of the
head, even when the direction-of-arrival of the desired speaker is
exactly known. In this paper, the head shadowing effect is taken
into account for CDR estimation by using a simplified model for the
frequency-dependent interaural time difference and a model for the
binaural coherence of the diffuse noise field. Evaluation of
CDR-based dereverberation with measured binaural impulse responses
indicates that the proposed binaural CDR estimators can improve PESQ
scores.
\end{abstract}
\section{Introduction}
\label{sec:intro}

Both speech quality and speech intelligibility may dramatically
degrade in reverberant and noisy environments. Many different
algorithms were proposed to suppress noise and the reverberation
during the past decades (see
\cite{IEEEhowto:Brandstein2001,IEEEhowto:Benesty2005,IEEEhowto:Naylor2010}
and references therein). This paper focuses on binaural speech
dereverberation, where the binaural signals are recorded with two
microphones located at two human ears.

Previous studies have already shown that it is important to preserve
both the interaural time difference (ITD) and the interaural level
difference (ILD) cues when applying binaural dereverberation methods
for hearing aids
\cite{IEEEhowto:Allen1977,IEEEhowto:Lebart1998,IEEEhowto:Jeub2010,IEEEhowto:Adam2014,IEEEhowto:Westermann2013,IEEEhowto:Tsilfidis2013},
since, when binaural cues are distorted, localization of sound
sources becomes difficult {\cite{IEEEhowto:Hamacher2005}}. This
condition is ensured by a two-channel postfiltering approach where
the same gain is applied to both channels \cite{IEEEhowto:Jeub2010}.
In {\cite{IEEEhowto:Jeub2010}}, Jeub \emph{et al.} took the
shadowing effect of the head into account in the diffuse sound field
model. In \cite{IEEEhowto:Westermann2013,IEEEhowto:Tsilfidis2013},
interaural coherence histograms were mapped to a gain function to
suppress the reverberant components in each frequency channel.


Recently, coherent-to-diffuse-ratio (CDR) estimators have been
proposed, which can be seen as an alternative formulation of
coherence-based dereverberation approaches \cite{IEEEhowto:Andreas2015}. In
{\cite{IEEEhowto:Jeub2010}}, the assumption was made that binaural
signals are time-aligned before calculating the spectral weights of
the Wiener filter. In {\cite{IEEEhowto:Thiergart2012}}, two CDR
estimators were proposed, where one requires knowledge on both the
direction of arrival (DOA) of the desired speaker and the spatial
coherence of the late reverberant speech, and the other does not
need the DOA information. In {\cite{IEEEhowto:Andreas2015}}, Schwarz
and Kellermann proposed improved estimators both for the case of
known and unknown DOA, which were shown to lead to improved
dereverberation performance (see \cite[Table
III]{IEEEhowto:Andreas2015} for details). To the best of our
knowledge, these CDR estimators have not been applied to binaural
dereverberation and their performance has not been reported until
now.

After briefly reviewing CDR estimators for free-field conditions, i.e., for a sound field with no obstructions close to the microphones,
we describe models for the ITD and the coherence of diffuse noise
under the influence of the head in a binaural scenario, and show
that the direction-dependent CDR estimators based on a free-field
assumption are not robust under this model. We propose to modify the
CDR estimators to use binaural models. Experimental
results confirm that the proposed estimators achieve higher PESQ
scores than the free-field estimators when applied to
coherence-based dereverberation. The proposed binaural CDR
estimators have numerous applications, such as binaural hearing
aids, robotics, or immersive audio communication systems.

\section{Free-field Signal Model and CDR Estimation}
We model two reverberant and noisy
microphone signals $x_i(t)$, $i=1,2$, as the sum of
a desired speech component $x_{i,{\rm coh}} (t)$  and an
undesired component $x_{i,{\rm diff}}
(t)$ consisting of diffuse
reverberation and/or noise:
\begin{equation}
\label{eq1} x_i (t) = x_{i,{\rm coh}} (t)
+ x_{i,{\rm diff}} (t),
\end{equation}
As in previous studies, we assume both microphones to be
omnidirectional and the desired component to be a plane wave in the
free (locally unobstructed) field, so that $x_{2,{\rm coh}} (t)$ is
a time-shifted version of $x_{1,{\rm coh}} (t)$
{\cite{IEEEhowto:Jeub2010,IEEEhowto:Thiergart2012,IEEEhowto:Andreas2015}}:
\begin{equation}
\label{eq2} x_{2,{\rm coh}} (t) = x_{1,{\rm coh}}
( {t - \tau _{12} } ),
\end{equation}
where $\tau _{12}$ is the time difference of arrival (TDOA) of the desired sound
between the first and the second microphone.
The free-field model for the spatial coherence between the desired speech component
at both microphones, $x_{1,{\rm coh}}(t)$ and $x_{2,{\rm coh}}(t)$,
is given by
\begin{equation}
\label{eq:planewave_ff} \Gamma^{\rm FF}_{\rm coh}(f) =
\exp(j\tau_{12}).
\end{equation}
If $\theta=0^{\circ}$ corresponds to broadside direction, the
TDOA in the free field can be
expressed as
\begin{equation}
\label{eq3} \tau _{12}  = d\sin \theta /c,
\end{equation}
where $d$ is the distance of the two microphones and $c$ is the
speed of sound.

The spatial coherence between the reverberation/noise components $x_{1,{\rm diff}}(t)$ and $x_{2,{\rm diff}}(t)$ is
given by the spatial coherence function of two omnidirectional
sensors in a diffuse (spherically isotropic), locally unobstructed sound field:
\begin{equation}
\label{eq:gamma_diff_ff} \Gamma^{\rm FF}_{{\rm diff}} \left( f \right) =
\frac{{\sin \left( {2\pi f d/c} \right)}}{{2\pi fd/c}},
\end{equation}
where $f$ is the frequency in Hz. For the cylindrically isotropic
field, the spatial coherence can be given by
\begin{equation}
\label{eq5} \Gamma^{\rm FF}_{{\rm 2D{\text{-}}iso}} \left( f \right) = J_0
\left( {2\pi fd/c} \right).
\end{equation}
Generally, (\ref{eq:gamma_diff_ff}) often fits better than
(\ref{eq5}) in practical applications \cite{IEEEhowto:Andreas2015}, therefore we use $\Gamma^{\rm
FF}_{{\rm diff}} \left( f \right)$ in this paper, although
$\Gamma^{\rm FF}_{{\rm 2D{\text{-}}iso}} \left( f \right)$ may be
applied analogously.

The CDR at the $i$-th
microphone can be given by
\begin{equation}
\mathit{CDR}_i(k,f)  = \frac{{\Phi _{i,{\rm coh}}(k,f)
}}{{\Phi _{i,{\rm diff}} (k,f)}},
\end{equation}
where $\Phi _{i,{\rm coh}} \left(k, f \right)$ and
$\Phi _{i,{\rm diff}} \left(k, f \right)$ are the short-time power
spectra of $x_{i,\rm coh}\left(t\right)$ and 
$x_{i,\rm diff}\left(t\right)$, respectively, with the frame index $k$ and frequency $f$ (we will omit both $k$ and $f$ in the following for brevity). We further assume that the power
spectra are identical at the two microphones for both the desired and undesired component, i.e., $\Phi _{{\rm coh}}  = \Phi
_{{1,\rm coh}} = \Phi _{{2, \rm coh}}$ and $\Phi _{{\rm diff}}  = \Phi _{{1,\rm
diff}}  = \Phi _{{2, \rm diff}}$, and therefore
\begin{equation}
\mathit{CDR} = \mathit{CDR}_1  = \mathit{CDR}_2  = \frac{{\Phi _{{\rm coh}} }}{{\Phi _{{\rm
diff}} }}.
\end{equation}

Using the models for the coherence of the desired and diffuse signal
components given above, and a short-time estimate of the coherence
between $x_1(t)$ and $x_2(t)$, which is in the following denoted as
$\hat\Gamma_x(k,f)$ and which may be obtained by recursive averaging,
it is possible to estimate the time- and frequency-dependent CDR, as described in detail in
\cite{IEEEhowto:Andreas2015}. The CDR estimators which are evaluated
in this paper are summarized in Table~\ref{tab:Table1}.

\begin{table}
\caption{Summary of CDR estimators evaluated in this paper.
$\Gamma_{\rm coh}$ and $\Gamma_{\rm diff}$ indicate the model
coherence functions used for desired signal and diffuse noise,
respectively, $\hat \Gamma_x$ indicates the estimated coherence of
the mixed sound field. $\Re\{\bullet\}$ extracts the real part of a
complex value and $^*$ denotes the complex conjugate.} \centering
\vspace{0mm} \footnotesize
\renewcommand{\arraystretch}{1.3}
\begin{tabular}{ll}
\hline
\textbf{Estimator}    &\textbf{Direction-dependent} \\
\hline
$\tilde \eta_{\rm Schwarz1}$ &${{\Re\left\{ {\Gamma _{{\rm coh}}^*
\left( {\Gamma _{{\rm diff}}  - \hat \Gamma _{x}} \right)} \right\}}
\mathord{\left/
 {\vphantom {{\Re\left\{ {\Gamma _{{\rm coh}}^* \left( {\Gamma _{{\rm diff}}  - \hat \Gamma_x } \right)} \right\}} {\left( {\Re\left\{ {\Gamma _{{\rm coh}}^* \hat \Gamma_x } \right\} - 1} \right)}}} \right.
 \kern-\nulldelimiterspace} {\left( {\Re\left\{ {\Gamma _{{\rm coh}}^* \hat \Gamma_x } \right\} - 1} \right)}}$\\
$\tilde \eta_{\rm Schwarz2}$ &$\left| {{{\Gamma _{{\rm coh}}^*
\left( {\Gamma _{{\rm diff}}  - \hat \Gamma _{x}} \right)}
\mathord{\left/
 {\vphantom {{\Gamma _{{\rm coh}}^* \left( {\Gamma _{{\rm diff}}  - \hat \Gamma_x } \right)} {\left( {\Re\left\{ {\Gamma _{{\rm coh}}^* \hat \Gamma_x } \right\} - 1} \right)}}} \right.
 \kern-\nulldelimiterspace} {\left( {\Re\left\{ {\Gamma _{{\rm coh}}^* \hat \Gamma_x } \right\} - 1} \right)}}} \right|$\\
\hline \textbf{Estimator}  & \textbf{Direction-independent} \\
\hline $\tilde \eta_{\rm Thiergart2}$ &$\Re\left\{ {{{\left( {\Gamma
_{{\rm diff}}  - \hat \Gamma_x } \right)}
\mathord{\left/
 {\vphantom {{\left( {\Gamma _{{\rm diff}}  - \hat \Gamma_x } \right)} {\left( {\hat \Gamma_x  - \exp \left( {j\angle \hat \Gamma_x } \right)} \right)}}} \right.
 \kern-\nulldelimiterspace} {\left( {\hat \Gamma_x  - \exp \left( {j\angle \hat \Gamma_x } \right)} \right)}}}
 \right\}$\\
 $\tilde \eta_{\rm Schwarz3}$ &\cite[(25)]{IEEEhowto:Andreas2015}\\
\hline
\end{tabular}
\label{tab:Table1}
\end{table}

\section{Binaural Signal Model}
When the two microphones are placed at the two ears, the ITD is the
propagation delay of the desired sound from the left ear to the
right ear and the ILD measures the power level difference between
the two microphones. Both the ITD and the ILD have already been
widely studied, and various models can be found in
{\cite{IEEEhowto:Blauert1997,IEEEhowto:Blauert2013}} and references
therein. As in {\cite{IEEEhowto:Jeub2010}}, the impact of the ILD is
neglected in the following, i.e., we maintain the assumption of
equal power at both microphones. Based on this
assumption, both the CDRs and the postfilter gain functions are the
same at the two microphones placed at the two ears.

In this section, we first describe a simplified model for the
frequency-dependent ITD and use it to derive a coherence model for
the desired signal component. Then, we describe appropriate models
for the diffuse sound field coherence which account for the effect
of the head. Finally, we describe the application of these models
for binaural CDR estimation and compare the robustness of CDR
estimators based on the free-field model to the binaural CDR
estimators.

\subsection{ITD and Desired Signal Coherence Model}

Previous studies have shown that, unlike the TDOA in the free-field case given by \eqref{eq3}, the ITD is highly
dependent on the frequency, the azimuth angle, the elevation angle
and the distance of the desired speaker from the head
\cite{IEEEhowto:TianshuQu2009,IEEEhowto:Blauert2013,IEEEhowto:Kuhn1977}.
Here, we use a simplified ITD model to make it
applicable for practical application to binaural dereverberation.
We assume that the distance of the desired speaker from the head is larger than 1m,
and thus does not have a significant effect on the ITD
{\cite[Fig.$\,$9]{IEEEhowto:TianshuQu2009}}. Furthermore, we neglect
separate consideration of elevation and azimuth angles, and instead
model the ITD as a function of the angle $\theta$, which we define as the angle
between the direction of the desired speaker and the
forward median plane of the head. According to the head-related
spherical coordinate system
{\cite[Fig.$\,$7]{IEEEhowto:TianshuQu2009}}, $\theta = 0$
and $\theta = \pm \pi$ correspond to the forward and the
backward median planes of the head, respectively.


Kuhn has shown that the ITD is frequency-dependent
{\cite{IEEEhowto:Kuhn1977}}, which can be approximately summarized
as
{
\begin{equation}
\label{eq7} \tau^L _{lr} ={\frac{{1.5d\sin \left( \theta  \right)}}{c}, {\rm for}\; f \le 500{\rm Hz}},  \\
\end{equation}}
and, for $f\ge f_H = 2000$, {
\begin{equation}
\label{eq8} \tau _{lr}^H = \left\{ {\begin{array}{*{20}l}
   {{{0.5d\left( {\sin \left( \theta  \right) + \theta } \right)} \mathord{\left/
 {\vphantom {{0.5d\left( {\sin \left( \theta  \right) + \theta } \right)} c}} \right.
 \kern-\nulldelimiterspace} c},\theta  \in \left[ {{\rm  - }{\pi }/{{\rm 2}},{\pi }/{{\rm 2}}} \right]}, \qquad\;\; \\
   {{{0.5d\left( {\sin \left( \theta  \right) - \left( {\pi  + \theta } \right)} \right)} \mathord{\left/
 {\vphantom {{0.5d\left( {\sin \left( \theta  \right) - \left( {\pi  + \theta } \right)} \right)} c}} \right.
 \kern-\nulldelimiterspace} c},\theta  \in \left[ {{\rm  - }\pi , - {\pi }/{{\rm 2}}} \right]},  \\
   {{{0.5d\left( {\sin \left( \theta  \right) + \left( {\pi  - \theta } \right)} \right)} \mathord{\left/
 {\vphantom {{0.5d\left( {\sin \left( \theta  \right) + \left( {\pi  - \theta } \right)} \right)} c}} \right.
 \kern-\nulldelimiterspace} c},\theta  \in \left[ {{\pi }/{{\rm 2}},\pi } \right]}, \quad\; \\
\end{array}} \right.
\end{equation}}
where (\ref{eq7}) and (\ref{eq8}) are identical to
{\cite[(7) and (12)]{IEEEhowto:Kuhn1977}}, respectively. However, for the middle
frequency range, there is not an explicit expression. We propose to use
a linear interpolation to model the ITD in the middle frequency
range, which agrees well with the measurement results {\cite{IEEEhowto:Kuhn1977}}
and is given by
\begin{equation}
\label{eq9}\tau _{lr}^{{\rm Mid}} \left( f \right) = \tau _{lr}^L
 + \frac{{\tau _{lr}^H  - \tau
_{lr}^L }}{{f_H  - f_L }}\left( {f - f_L } \right),
\end{equation}
where $f = f_{\rm Mid} \in [500\;\;2000]$Hz.

\begin{figure}
  \centering
  \includegraphics[width=7cm]{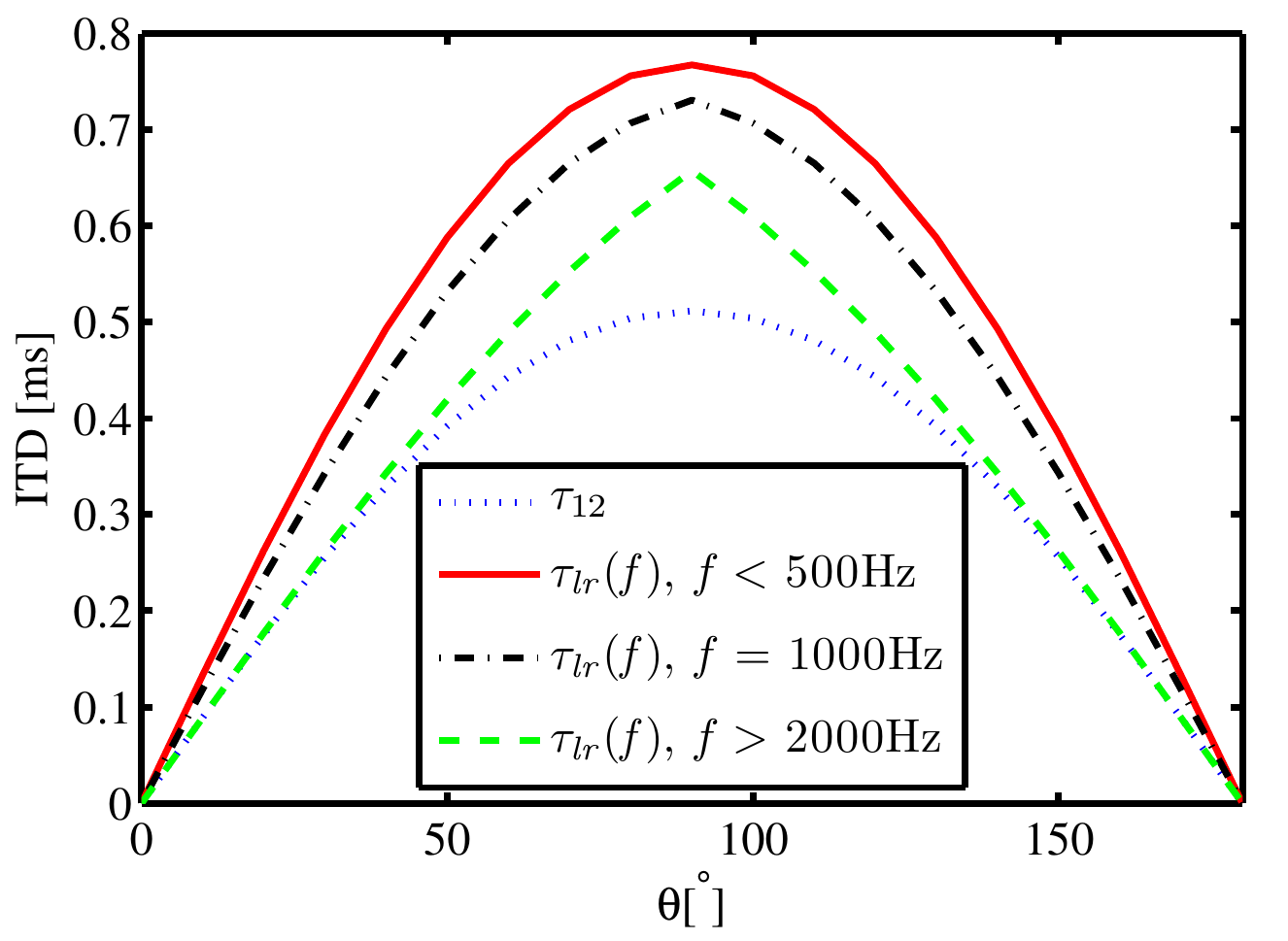}
  \vspace{0mm}
\caption{Comparison of $\tau_{12}$ and $\tau_{lr}(f)$ versus the
angle of the desired sound for different frequencies.} \label{fig1}
\end{figure}

Compared to $\tau_{12}$, the ITD $\tau_{lr}(f)$ is not only a
function of the DOA but also of the frequency. $\tau_{12}$ and
$\tau_{lr}(f)$ versus $\theta$ are plotted in Fig. {\ref{fig1}} for
different frequencies. Fig. {\ref{fig1}} shows that the difference
between $|\tau_{lr}(f)|$ and $|\tau_{12}|$ is largest for $f \le
500$Hz. For $f>2000$Hz, $\tau_{lr}(f)$ is close to $\tau_{12}$ when
$|\theta|$ or $|\pi\pm\theta|$ is smaller than $\pi/4$, while
$|\tau_{lr}(f)|$ is much larger than $|\tau_{12}|$ for $|\theta|$
close to $\pi/2$.

Without the shadowing effect of the head, the free-field coherence model
of the desired signal is given by (\ref{eq:planewave_ff}). Based on the
frequency-dependent ITD model which accounts for the head effect, we can now define the coherence of the desired component
for the binaural case as:
\begin{equation}
\label{eq11} \Gamma^{\rm Binaural}_{\rm coh} \left( f \right) = \exp \left(
{j\tau _{lr} \left( f \right)} \right).
\end{equation}


\subsection{Diffuse Noise Coherence Model}
The shadowing effect of the head also has an impact on the spatial
coherence of the two microphone signals in a diffuse sound field.
Both theoretical results and experimental results can be found in
{\cite{IEEEhowto:Lindevald1986,IEEEhowto:JeubSPL2011}}. Here we use
the analytic representation of the binaural correlation function
proposed by Lindevald and Benade {\cite{IEEEhowto:Lindevald1986}},
given by
\begin{equation}
\label{eq:lindevald} \Gamma^{\rm Binaural}_{\rm diff} \left( f \right) =
\frac{1}{{ \sqrt{1 + \left( {\beta 2\pi f{d \mathord{\left/
 {\vphantom {d c}} \right.
 \kern-\nulldelimiterspace} c}} \right)^4 } }}\frac{{\sin \left( {\alpha 2\pi f{d \mathord{\left/
 {\vphantom {d c}} \right.
 \kern-\nulldelimiterspace} c}} \right)}}{{\left( {\alpha 2\pi f{d \mathord{\left/
 {\vphantom {d c}} \right.
 \kern-\nulldelimiterspace} c}} \right)}},
\end{equation}
where $\alpha=2.2$ and $\beta=0.5$.

The binaural CDR estimators are now obtained by inserting the
binaural coherence models $\Gamma^{\rm Binaural}_{\rm coh}$ and
$\Gamma^{\rm Binaural}_{\rm diff}$ into the estimators given in
Table {\ref{tab:Table1}}. This extension makes the
direction-dependent CDR estimators suitable for binaural
dereverberation. The corresponding estimators are denoted as $\tilde
\eta^{\rm Binaural}_{\bullet}$ in the following,
where ${\bullet}$ represents the name of the
technique that is being used.


\begin{figure}
  \centering
  \includegraphics[width=7cm]{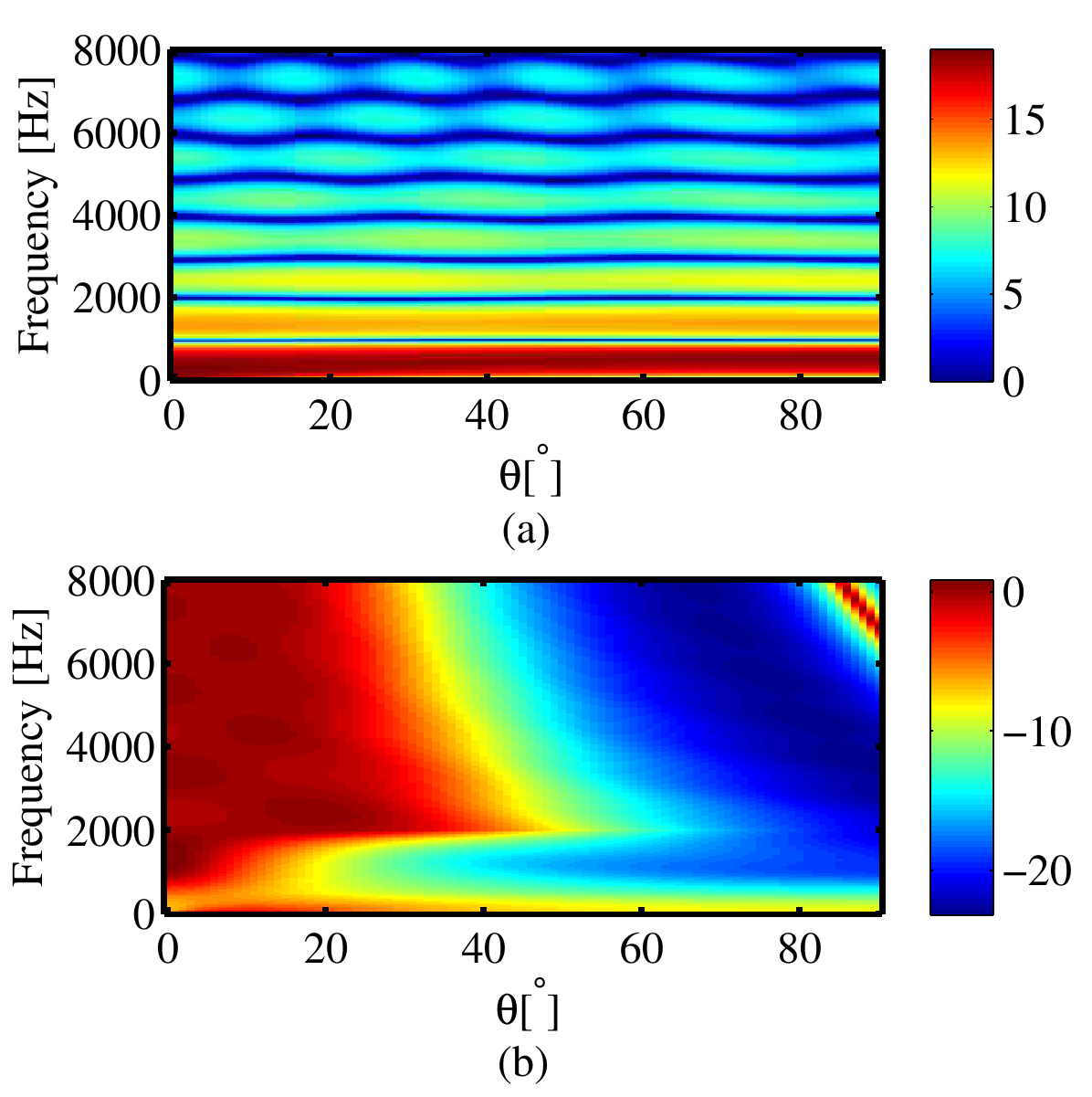}
  \vspace{-2mm}
\caption{CDR estimation error of the free-field estimator, $\Delta$, versus $f$ and $\theta$
for (a) the input CDR $\eta_{\rm in} = -20$dB, (b) the input CDR
$\eta_{\rm in} = 20$dB.} \label{fig3}
\end{figure}

\begin{table*}
\caption{PESQ scores averaged over all angles for CDR estimators in
Table {\ref{tab:Table1}}, using free-field ($\tilde\eta^{\rm
FF}_{\bullet}$) or binaural coherence models ($\tilde\eta^{\rm
Binaural}_{\bullet}$).} \footnotesize \centering

\renewcommand{\arraystretch}{1.3}
\vspace{0mm}
\begin{tabular}{c|c|cc|cc|cc|cc}
\hline {AIR} &{Unprocessed}
&\multicolumn{4}{c|}{Direction-dependent}
&\multicolumn{4}{c}{Direction-independent} \\ \hline
\textbf{Distance} &Left/Right &$\tilde\eta^{\rm FF}_{\rm Schwarz1}$
&$\tilde\eta^{\rm Binaural}_{\rm Schwarz1}$ &$\tilde\eta^{\rm FF}_{\rm Schwarz2}$
&$\tilde\eta^{\rm Binaural}_{\rm Schwarz2}$ &$\tilde\eta^{\rm FF}_{\rm
Thiergart2}$ &$\tilde\eta^{\rm Binaural}_{\rm Thiergart2}$
&$\tilde\eta^{\rm FF}_{\rm Schwarz3}$
&$\tilde\eta^{\rm Binaural}_{\rm Schwarz3}$ \\
\hline 1\,m &2.24/2.25 &2.40/2.41 &2.65/2.68 &2.57/2.59
&\textbf{2.69/2.71} &2.66/2.67 &2.64/2.65 &2.65/2.67 &2.64/2.65 \\
2\,m &1.88/1.90 &2.00/2.00 &2.12/2.13 &2.10/2.10
&\textbf{2.17/2.18}
&2.16/2.17 &2.15/2.15 &2.16/2.17 &2.15/2.16 \\
3\,m &1.77/1.77 &1.85/1.84 &1.91/1.90 &1.92/1.91
&\textbf{1.97/1.96}
&1.95/1.95 &1.95/1.95 &1.96/1.96 &1.95/1.95 \\
\hline
\end{tabular}
\label{tab:Table3}
\end{table*}

\subsection{Robustness of the Free-Field Estimators
in the Binaural Scenario} This part evaluates the robustness of the
direction-dependent CDR estimators using the free-field model
against the shadowing effect of the head. For the limited space of
this paper, only $\tilde\eta^{\rm Binaural}_{\rm Schwarz2}$ is
chosen to compare with $\tilde\eta^{\rm FF}_{\rm Schwarz2}$, since a
previous study {\cite{IEEEhowto:Andreas2015}} has already shown that
$\tilde\eta^{\rm FF}_{\rm Schwarz2}$ has the best performance among
the direction-dependent CDR estimators in Table {\ref{tab:Table1}}
(see {\cite[Table III]{IEEEhowto:Andreas2015}} for details). For the
comparison, we generate values of the mixture coherence
$\hat\Gamma_x$ for a certain input CDR $\eta_{\rm in}$ and different
angles and frequencies according to the binaural coherence models
defined above, and insert these coherence values into the free-field
estimator. We then define the estimation error of the free-field CDR
estimator compared to the true CDR $\eta_{\rm in}$ as
\begin{equation}
\label{eq12}\Delta  = 10\log _{10} \tilde\eta^{\rm FF}_{{\rm
Schwarz2}} - 10\log _{10} \eta_{\rm in}.
\end{equation}

Fig. {\ref{fig3}} plots $\Delta$ versus $f$ and $\theta$ for the
true input CDR $\eta_{\rm in} = -20$dB (a) and $\eta_{\rm in} =
20$dB (b). Only $\theta \in [0\;\pi/2]$ is considered due to the
symmetry of the scenario. Fig. {\ref{fig3}} shows that the CDR is
somewhat overestimated for low input CDR, while for high input CDR,
the CDR is seriously underestimated for angles larger than
45$^\circ$. The influence of the head on the coherence, especially
the one of the desired speech component $\Gamma^{\rm Binaural}_{{\rm
coh}}(f)$, is significant enough to deteriorate the performance of
the free-field CDR estimator considerably.

\section{Evaluation}
This section evaluates the application of the CDR estimators in
Table \ref{tab:Table1} with the free-field and binaural coherence
models to the problem of dereverberation. We use the Aachen Impulse
Response (AIR) database {\cite{IEEEhowto:Jeub2009}}, which consists
of binaural RIRs measured by a dummy head with azimuth angles from
-90$^{\circ}$ to 90$^{\circ}$ with 15$^{\circ}$ increments and
source-head distances from 1\,m to 3\,m with 1\,m increments.


Ten clean speech samples (five female and five male speakers) are
taken from the TIMIT database {\cite{IEEEhowto:TIMIT1993}}. The
reverberant speech samples are generated by convolving the clean
speech with the ``stairway'' RIRs from the AIR database. We use the
same filterbank,  postfilter gain function and parameters as in
{\cite[(29)]{IEEEhowto:Andreas2015}}, with the CDR estimators in
Table \ref{tab:Table1}. Knowledge of the true DOA is assumed for
computation of the desired signal coherence models. The gain
function is applied to the two microphone signals separately. PESQ
is chosen as evaluation measure since it was found to be highly correlated with speech quality
for the evaluation of noise and reverberation suppression methods
{\cite{IEEEhowto:HuLoizou2008,Goetze2014}}. Here, we give raw MOS scores
obtained by wideband PESQ. The
PESQ scores of the two microphone signals and those of the processed
signals are given separately. Note that the average PESQ scores for
both ears are very similar, due to the symmetry of the scenario. The
experimental results for the different distances are presented in
Table {\ref{tab:Table3}}.
From these results, we can make the
following observations:
\begin{enumerate}
\item[(1)] Using the ITD and binaural diffuse coherence model can improve all of the
direction-dependent CDR estimators.
\item[(2)] The direction-independent CDR estimators, which
do not rely on a model of the desired signal coherence, are robust in the binaural case, even when using the free-field diffuse coherence model. This indicates that the choice of the diffuse coherence model is not critical, and the main effect of the head is on the ITD.
\item[(3)] The direction-independent estimators almost reach the performance of the best binaural direction-dependent estimator.
\end{enumerate}

\begin{figure}[t!]
  \centering
  \includegraphics[width=\columnwidth]{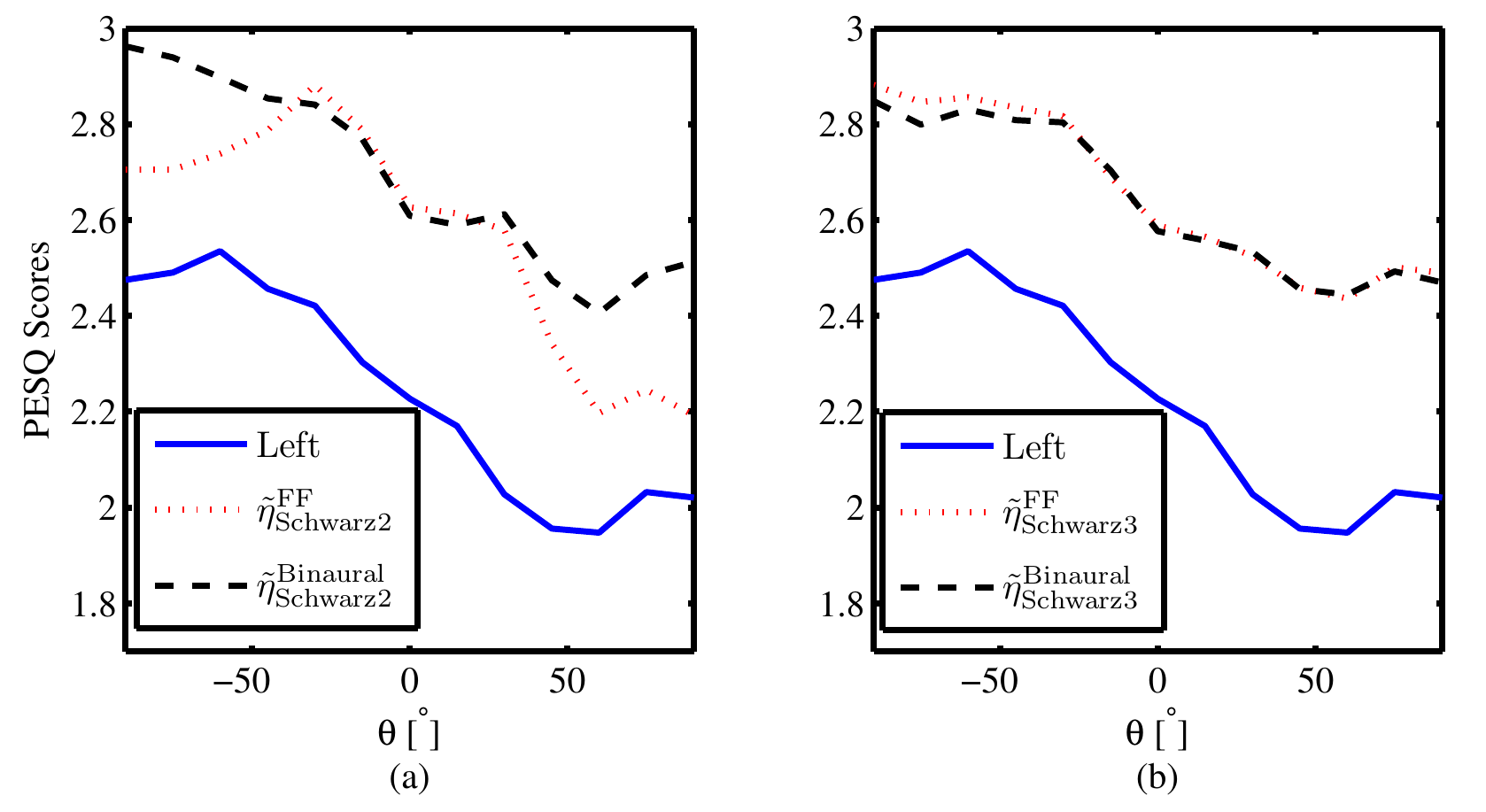}
  \vspace{-5mm}
\caption{PESQ scores versus DOA for the 1\,m distance case: (a)
direction-dependent CDR estimators; (b) direction-independent CDR
estimators. \textit{Left} represents the unprocessed signal recorded
by the microphone located at the left ear.} \label{fig4}
\end{figure}

To reveal the mechanism of the better performance of the proposed
direction-dependent binaural CDR estimators, the PESQ scores versus
$\theta$ are plotted in Fig. {\ref{fig4}} (due to
symmetry, only PESQ scores for the left microphone are shown). As can be seen from this figure,
$\tilde\eta^{\rm Binaural}_{\rm Schwarz2}$ is much better than
$\tilde\eta^{\rm FF}_{\rm Schwarz2}$ for $|\theta| \ge 45^{\circ}$.
This phenomenon can be explained by the robustness analysis results
in Fig.~{\ref{fig3}}, where it was found that the estimation error of
$\tilde\eta^{\rm
FF}_{\rm Schwarz2}$ becomes significant for $|\theta| > 45^{\circ}$. However,
$\tilde\eta^{\rm FF}_{\rm Schwarz3}$ and $\tilde\eta^{\rm
Binaural}_{\rm Schwarz3}$ nearly have the same performance for all
angles, which confirms that the effect of using $\Gamma^{\rm FF}_{\rm diff}\left(
f \right)$ or $\Gamma^{\rm Binaural}_{\rm diff}\left( f \right)$ is
not critical for the direction-independent CDR estimators.

The estimators $\tilde\eta_{\rm Thiergart2}$ and
$\tilde\eta_{\rm Schwarz3}$ show similar behavior in this scenario,
although the former is biased \cite{IEEEhowto:Andreas2015}.
This can be explained by the fact that the bias is roughly proportional to the noise coherence and disappears for $\Gamma_{\rm diff} \to 0$; since, for binaural signals, the noise coherence is lower than for the setup investigated in \cite{IEEEhowto:Andreas2015}, due to the large spacing of the sensors and the shadowing effect of the head, the practical impact of the bias is not significant here.

\section{Conclusions}
This paper extends previously proposed free-field CDR estimators to
binaural dereverberation by using a simplified model for the ITD.
Experimental results show that this extension is important for the
direction-dependent CDR estimators, where PESQ scores for
dereverberation can be significantly improved. It is further shown
that the direction-independent CDR estimators, which do not
require a model of the desired signal coherence, can achieve similar
performance and are robust towards the shadowing effect
of the head. Further work could concentrate on studying the impact
of the ILD on binaural dereverberation and the theoretical limits of
the CDR estimators by using statistical analysis
{\cite{IEEEhowto:Zheng2013}}.
%


\end{document}